\documentstyle[12pt]{article}
\def\doit#1#2{\ifcase#1\or#2\fi}
\skewchar\fivmi='177
\skewchar\sixmi='177
\skewchar\sevmi='177
\skewchar\egtmi='177
\skewchar\ninmi='177
\skewchar\tenmi='177
\skewchar\elvmi='177
\skewchar\twlmi='177
\skewchar\frtnmi='177
\skewchar\svtnmi='177
\skewchar\twtymi='177
\def\@magscale#1{ scaled \magstep #1}

\def\framingfonts#1{
\doit{#1}{\font\twfvmi  = ammi10   \@magscale5 
\skewchar\twfvmi='177
\skewchar\fivsy='60
\skewchar\sixsy='60
\skewchar\sevsy='60
\skewchar\egtsy='60
\skewchar\ninsy='60
\skewchar\tensy='60
\skewchar\elvsy='60
\skewchar\twlsy='60
\skewchar\frtnsy='60
\skewchar\svtnsy='60
\skewchar\twtysy='60
\font\twfvsy  = amsy10   \@magscale5 
\skewchar\twfvsy='60
\font\go=font018			
\font\sc=font005			
\def\Go#1{{\hbox{\go #1}}}	
\def\Sc#1{{\hbox{\sc #1}}}	
\def\Sf#1{{\hbox{\sf #1}}}	
\font\oo=circlew10	      
\font\ooo=circle10			
\font\ro=manfnt				
\def\kcl{{\hbox{\ro 6}}}		
\def\kcr{{\hbox{\ro 7}}}		
\def\ktl{{\hbox{\ro \char'134}}}	
\def\ktr{{\hbox{\ro \char'135}}}	
\def\kbl{{\hbox{\ro \char'136}}}	
\def\kbr{{\hbox{\ro \char'137}}}	
}}

\catcode`@=11
\def\un#1{\relax\ifmmode\@@underline#1\else
	$\@@underline{\hbox{#1}}$\relax\fi}
\catcode`@=12

\let\du=\d			

\def\a{\alpha}
\def\b{\beta}

\def\d{\delta}
\def\e{\epsilon}

\def\k{\kappa}

\def\m{\mu}
\def\n{\nu}
\def\o{\omega}

\def\r{\rho}
\def\s{\sigma}
\def\t{\tau}

\def\z{\zeta}

\def\L{\Lambda}

\def\S{\Sigma}

\def\plpl{{\raise-2pt\hbox{$\raise3pt\hbox{$_+$}\hskip-7.0pt\raise0.0pt
\hbox{$^+$}\hskip 0.01pt$}}}
\def\mimi{{\raise-2pt\hbox{$\raise3pt\hbox{$_-$}\hskip-7.0pt\raise0.0pt
\hbox{$^-$}\hskip 0.01pt$}}}

\def\bo{{\raise.15ex\hbox{\large$\Box$}}}		
\def\pr{\prod}						
\def\TH{{\raise.2ex\hbox{$\displaystyle \bigodot$}\mskip-4.7mu \llap H \;}}
\def\face{{\raise.2ex\hbox{$\displaystyle \bigodot$}\mskip-2.2mu \llap {$\ddot
	\smile$}}}					

\def\sp#1{{}^{#1}}				
\def\Tilde#1{{\widetilde{#1}}\hskip 0.03in}			
\def\Hat#1{\widehat{#1}}			
\def\leftrightarrowfill{$\mathsurround=0pt \mathord\leftarrow \mkern-6mu
	\cleaders\hbox{$\mkern-2mu \mathord- \mkern-2mu$}\hfill
	\mkern-6mu \mathord\rightarrow$}
\def\dvec#1{\vbox{\ialign{##\crcr
	\leftrightarrowfill\crcr\noalign{\kern-1pt\nointerlineskip}
	$\hfil\displaystyle{#1}\hfil$\crcr}}}		

\def\frac#1#2{{\textstyle{#1\over\vphantom2\smash{\raise.20ex
	\hbox{$\scriptstyle{#2}$}}}}}			
\def\sfrac#1#2{{\vphantom1\smash{\lower.5ex\hbox{\small$#1$}}\over
	\vphantom1\smash{\raise.4ex\hbox{\small$#2$}}}}	
\def\bfrac#1#2{{\vphantom1\smash{\lower.5ex\hbox{$#1$}}\over
	\vphantom1\smash{\raise.3ex\hbox{$#2$}}}}	
\def\afrac#1#2{{\vphantom1\smash{\lower.5ex\hbox{$#1$}}\over#2}}    

\newskip\humongous \humongous=0pt plus 1000pt minus 1000pt
\def\caja{\mathsurround=0pt}
\def\eqalign#1{\,\vcenter{\openup2\jot \caja
	\ialign{\strut \hfil$\displaystyle{##}$&$
	\displaystyle{{}##}$\hfil\crcr#1\crcr}}\,}
\newif\ifdtup
\def\panorama{\global\dtuptrue \openup2\jot \caja
	\everycr{\noalign{\ifdtup \global\dtupfalse
	\vskip-\lineskiplimit \vskip\normallineskiplimit
	\else \penalty\interdisplaylinepenalty \fi}}}
\def\li#1{\panorama \tabskip=\humongous				
	\halign to\displaywidth{\hfil$\displaystyle{##}$
	\tabskip=0pt&$\displaystyle{{}##}$\hfil
	\tabskip=\humongous&\llap{$##$}\tabskip=0pt
	\crcr#1\crcr}}

\doit0{
\def\ref#1{$\sp{#1)}$}
}

\parskip=\medskipamount			
\thicklines			    

\def\oldheadpic{				
	\setlength{\unitlength}{.4mm}
	\thinlines
	\par
	\begin{picture}(349,16)
	\put(325,16){\line(1,0){4}}
	\put(330,16){\line(1,0){4}}
	\put(340,16){\line(1,0){4}}
	\put(335,0){\line(1,0){4}}
	\put(340,0){\line(1,0){4}}
	\put(345,0){\line(1,0){4}}
	\put(329,0){\line(0,1){16}}
	\put(330,0){\line(0,1){16}}
	\put(339,0){\line(0,1){16}}
	\put(340,0){\line(0,1){16}}
	\put(344,0){\line(0,1){16}}
	\put(345,0){\line(0,1){16}}
	\put(329,16){\oval(8,32)[bl]}
	\put(330,16){\oval(8,32)[br]}
	\put(339,0){\oval(8,32)[tl]}
	\put(345,0){\oval(8,32)[tr]}
	\end{picture}
	\par
	\thicklines
	\vskip.2in}
\def\oldtitle#1#2#3#4{\oldheadpic\begin{center}\vglue.5in{\large\bf #1}\\[.6in]
	{#2}\\[.1in] {\it Department of Physics and Astronomy}\\
	{\it University of Maryland, College Park, MD 20742}\\[.6in]
	Physics Publication \#{#3}\\ {#4}\\[1.5in] {\bf Abstract}\\[.1in]
	\end{center} \begin{quotation}}			
\def\oldTitle#1#2#3#4#5#6#7{\oldheadpic\begin{center} \vglue .4in
	{\large\bf #1}\\[.4in]
	{#2}\\[.1in] {\it Department of Physics and Astronomy}\\
	{\it University of Maryland, College Park, MD 20742}\\[.1in]
	{#3}\\[.1in] {\it {#4}}\\ {\it {#5}}\\[.4in]
	Physics Publication \#{#6}\\ {#7}\\[.5in] {\bf Abstract}\\[.1in]
	\end{center} \begin{quotation}}			
\def\border{						
	\setlength{\unitlength}{1mm}
	\newcount\xco
	\newcount\yco
	\xco=-24
	\yco=12
	\begin{picture}(140,0)
	\put(\xco,\yco){$\ktl$}
	\advance\yco by-1
	{\loop
	\put(\xco,\yco){$\kcl$}
	\advance\yco by-2
	\ifnum\yco>-240
	\repeat
	\put(\xco,\yco){$\kbl$}}
	\xco=158
	\yco=12
	\put(\xco,\yco){$\ktr$}
	\advance\yco by-1
	{\loop
	\put(\xco,\yco){$\kcr$}
	\advance\yco by-2
	\ifnum\yco>-240
	\repeat
	\put(\xco,\yco){$\kbr$}}
        \put(-20,11){\tiny University of Maryland Elementary Particle
Physics University of Maryland Elementary Particle Physics University of
Maryland Elementary Particle Physics}
	\put(-20,-241.5){\tiny University of Maryland Elementary
Particle Physics University of Maryland Elementary Particle Physics
University of Maryland Elementary Particle Physics}
	\end{picture}
	\par\vskip-8mm}
\def\bordero{						
	\setlength{\unitlength}{1mm}
	\newcount\xco
	\newcount\yco
	\xco=-24
	\yco=12
	\begin{picture}(140,0)
	\put(\xco,\yco){$\ktl$}
	\advance\yco by-1
	{\loop
	\put(\xco,\yco){$\kcl$}
	\advance\yco by-2
	\ifnum\yco>-240
	\repeat
	\put(\xco,\yco){$\kbl$}}
	\xco=158
	\yco=12
	\put(\xco,\yco){$\ktr$}
	\advance\yco by-1
	{\loop
	\put(\xco,\yco){$\kcr$}
	\advance\yco by-2
	\ifnum\yco>-240
	\repeat
	\put(\xco,\yco){$\kbr$}}
	\put(-20,12){\ooo
bacdefghidfghghdhededbihdgdfdfhhdheidhdhebaaahjhhdahbahgdedgehgfdiehhgdigicba}
	\put(-20,-241.5){\ooo
ababaighefdbfghgeahgdfgafagihdidihiidhiagfedhadbfdecdcdfagdcbhaddhbgfchbgfdacfediacbabab}
	\end{picture}
	\par\vskip-8mm}
\def\headpic{						
	\indent
	\setlength{\unitlength}{.4mm}
	\thinlines
	\par
	\begin{picture}(29,16)
	\put(165,16){\line(1,0){4}}
	\put(170,16){\line(1,0){4}}
	\put(180,16){\line(1,0){4}}
	\put(175,0){\line(1,0){4}}
	\put(180,0){\line(1,0){4}}
	\put(185,0){\line(1,0){4}}
	\put(169,0){\line(0,1){16}}
	\put(170,0){\line(0,1){16}}
	\put(179,0){\line(0,1){16}}
	\put(180,0){\line(0,1){16}}
	\put(184,0){\line(0,1){16}}
	\put(185,0){\line(0,1){16}}
	\put(169,16){\oval(8,32)[bl]}
	\put(170,16){\oval(8,32)[br]}
	\put(179,0){\oval(8,32)[tl]}
	\put(185,0){\oval(8,32)[tr]}
	\end{picture}
	\par\vskip-6.5mm
	\thicklines}
\def\title#1#2#3#4{\border\headpic {\hbox to\hsize{#4 \hfill UMDEPP #3}}\par
	\begin{center} \vglue .5in {\large\bf #1}\\[.6in]
	{#2}\\[.1in] {\it Department of Physics and Astronomy}\\
	{\it University of Maryland, College Park, MD 20742}\\[1.5in]
	{\bf Abstract}\\[.1in] \end{center} \begin{quotation}}	
\def\Title#1#2#3#4#5#6#7{\border\headpic
	{\hbox to\hsize{#7 \hfill UMDEPP #6}}\par
	\begin{center} \vglue .4in {\large\bf #1}\\[.4in]
	{#2}\\[.1in] {\it Department of Physics and Astronomy}\\
	{\it University of Maryland, College Park, MD 20742}\\[.1in]
	{#3}\\[.1in] {\it {#4}}\\ {\it {#5}}\\[.5in] {\bf Abstract}\\[.1in]
	\end{center} \begin{quotation}}			
\def\endtitle{\end{quotation}\newpage}			

\def\sect#1{\bigskip\medskip \goodbreak \noindent{\bf {#1}} \nobreak \medskip}
\def\refs{\sect{References} \footnotesize \frenchspacing \parskip=0pt}
\def\Item{\par\hang\textindent}

\def\doit#1#2{\ifcase#1\or#2\fi}
\def\[{\lfloor{\hskip 0.35pt}\!\!\!\lceil}
\def\]{\rfloor{\hskip 0.35pt}\!\!\!\rceil}

\def\du#1#2{_{#1}{}^{#2}}
\def\ud#1#2{^{#1}{}_{#2}}

\def\E{{\cal E}}
\def\Re{{\cal R}e\,}
\def\Im{{\cal I}m\,}
\def\order#1#2{{\cal O}({#1}^{#2})}

\def\pl#1#2#3{Phys.~Lett.~{\bf {#1}B} (19{#2}) #3}
\def\np#1#2#3{Nucl.~Phys.~{\bf B{#1}} (19{#2}) #3}
\def\prl#1#2#3{Phys.~Rev.~Lett.~{\bf #1} (19{#2}) #3}
\def\pr#1#2#3{Phys.~Rev.~{\bf D{#1}} (19{#2}) #3}

\def\jmp#1#2#3{Jour.~Math.~Phys.~{\bf {#1}} (19{#2}) #3}

\def\ptp#1#2#3{Prog.~Theor.~Phys.~{\bf {#1}} (19{#2}) #3}

\def\ibid#1#2#3{{\it ibid.}~{\bf {#1}} (19{#2}) #3}
\def\grg#1#2#3{Gen.~Rel.~Grav.~{\bf{#1}} (19{#2}) {#3} }
\def\pla#1#2#3{Phys.~Lett.~{\bf A{#1}} (19{#2}) {#3}}
\def\mpl#1#2#3{Mod.~Phys.~Lett.~{\bf A{#1}} (19{#2}) #3}

\def\rmp#1#2#3{Rev.~Mod.~Phys.~{\bf {#1}} (19{#2}) {#3}}

\def\fracmm#1#2{{{#1}\over{#2}}}
\def\half{{\fracm12}}

\def\frac#1#2{{\textstyle{#1\over\vphantom2\smash{\raise -.20ex
	\hbox{$\scriptstyle{#2}$}}}}}			
\def\fracm#1#2{\hbox{\large{${\frac{{#1}}{{#2}}}$}}}

\def\uln{{\underline n}}
\def\Tilde#1{{\widetilde{#1}}\hskip 0.015in}
\def\Hat#1{\widehat{#1}}

\def\scst{\scriptstyle}

\def\.{.$\,$}

\def\uln#1{\underline{#1}}

\def\ul{\underline}
\def\un{\underline}
\def\-{{\hskip 1.5pt}\hbox{-}}

\def\kd#1#2{\d\du{#1}{#2}}
\def\fracmm#1#2{{{#1}\over{#2}}}
\def\footnotew#1{\footnote{\hsize=6.5in {#1}}}

\def\low#1{{\raise -3pt\hbox{${\hskip 1.0pt}\!_{#1}$}}}

\begin{document}

\font\tenmib=cmmib10
\font\sevenmib=cmmib10 at 7pt 
\font\fivemib=cmmib10 at 5pt  
\font\tenbsy=cmbsy10
\font\sevenbsy=cmbsy10 at 7pt 
\font\fivebsy=cmbsy10 at 5pt  
\def\BMfont{\textfont0\tenbf \scriptfont0\sevenbf
                              \scriptscriptfont0\fivebf
            \textfont1\tenmib \scriptfont1\sevenmib
                               \scriptscriptfont1\fivemib
            \textfont2\tenbsy \scriptfont2\sevenbsy
                               \scriptscriptfont2\fivebsy}
\def\rlx{\relax\leavevmode}
\def\BM#1{\rlx\ifmmode\mathchoice
                      {\hbox{$\BMfont#1$}}
                      {\hbox{$\BMfont#1$}}
                      {\hbox{$\scriptstyle\BMfont#1$}}
                      {\hbox{$\scriptscriptstyle\BMfont#1$}}
                 \else{$\BMfont#1$}\fi}

\font\tenmib=cmmib10
\font\sevenmib=cmmib10 at 7pt 
\font\fivemib=cmmib10 at 5pt  
\font\tenbsy=cmbsy10
\font\sevenbsy=cmbsy10 at 7pt 
\font\fivebsy=cmbsy10 at 5pt  
\def\BMfont{\textfont0\tenbf \scriptfont0\sevenbf
                              \scriptscriptfont0\fivebf
            \textfont1\tenmib \scriptfont1\sevenmib
                               \scriptscriptfont1\fivemib
            \textfont2\tenbsy \scriptfont2\sevenbsy
                               \scriptscriptfont2\fivebsy}
\def\BM#1{\rlx\ifmmode\mathchoice
                      {\hbox{$\BMfont#1$}}
                      {\hbox{$\BMfont#1$}}
                      {\hbox{$\scriptstyle\BMfont#1$}}
                      {\hbox{$\scriptscriptstyle\BMfont#1$}}
                 \else{$\BMfont#1$}\fi}

\def\inbar{\vrule height1.5ex width.4pt depth0pt}
\def\sinbar{\vrule height1ex width.35pt depth0pt}
\def\ssinbar{\vrule height.7ex width.3pt depth0pt}
\font\cmss=cmss10
\font\cmsss=cmss10 at 7pt
\def\ZZ{\rlx\leavevmode
             \ifmmode\mathchoice
                    {\hbox{\cmss Z\kern-.4em Z}}
                    {\hbox{\cmss Z\kern-.4em Z}}
                    {\lower.9pt\hbox{\cmsss Z\kern-.36em Z}}
                    {\lower1.2pt\hbox{\cmsss Z\kern-.36em Z}}
               \else{\cmss Z\kern-.4em Z}\fi}
\def\Ik{\rlx{\rm I\kern-.18em k}}  
\def\IC{\rlx\leavevmode
             \ifmmode\mathchoice
                    {\hbox{\kern.33em\inbar\kern-.3em{\rm C}}}
                    {\hbox{\kern.33em\inbar\kern-.3em{\rm C}}}
                    {\hbox{\kern.28em\sinbar\kern-.25em{\rm C}}}
                    {\hbox{\kern.25em\ssinbar\kern-.22em{\rm C}}}
             \else{\hbox{\kern.3em\inbar\kern-.3em{\rm C}}}\fi}
\def\IP{\rlx{\rm I\kern-.18em P}}
\def\IR{\rlx{\rm I\kern-.18em R}}
\def\IN{\rlx{\rm I\kern-.20em N}}
\def\Ione{\rlx{\rm 1\kern-2.7pt l}}

\def\unredoffs{} \def\redoffs{\voffset=-.31truein\hoffset=-.59truein}
\def\speclscape{\special{ps: landscape}}

\newbox\leftpage \newdimen\fullhsize \newdimen\hstitle \newdimen\hsbody
\tolerance=1000\hfuzz=2pt\def\fontflag{cm}
\catcode`\@=11 
\doit0
{
\def\bigans{b }
\message{ big or little (b/l)? }\read-1 to\answ
\ifx\answ\bigans\message{(This will come out unreduced.}
}
\hsbody=\hsize \hstitle=\hsize 
\doit0{
\else\message{(This will be reduced.} \let\l@r=L
\redoffs \hstitle=8truein\hsbody=4.75truein\fullhsize=10truein\hsize=\hsbody
\output={\ifnum\pageno=0 
  \shipout\vbox{\speclscape{\hsize\fullhsize\makeheadline}
    \hbox to \fullhsize{\hfill\pagebody\hfill}}\advancepageno
  \else
  \almostshipout{\leftline{\vbox{\pagebody\makefootline}}}\advancepageno
  \fi}
}
\def\almostshipout#1{\if L\l@r \count1=1 \message{[\the\count0.\the\count1]}
      \global\setbox\leftpage=#1 \global\let\l@r=R
 \else \count1=2
  \shipout\vbox{\speclscape{\hsize\fullhsize\makeheadline}
      \hbox to\fullhsize{\box\leftpage\hfil#1}}  \global\let\l@r=L\fi}
\fi
\def\nolabels{\def\wrlabeL##1{}\def\eqlabeL##1{}\def\reflabeL##1{}}
\def\writelabels{\def\wrlabeL##1{\leavevmode\vadjust{\rlap{\smash%
{\line{{\escapechar=` \hfill\rlap{\sevenrm\hskip.03in\string##1}}}}}}}%
\def\eqlabeL##1{{\escapechar-1\rlap{\sevenrm\hskip.05in\string##1}}}%
\def\reflabeL##1{\noexpand\llap{\noexpand\sevenrm\string\string\string##1}}}
\nolabels
%
\global\newcount\secno \global\secno=0
\global\newcount\meqno \global\meqno=1
\def\newsec#1{\global\advance\secno by1\message{(\the\secno. #1)}
\global\subsecno=0\eqnres@t\noindent{\bf\the\secno. #1}
\writetoca{{\secsym} {#1}}\par\nobreak\medskip\nobreak}
\def\eqnres@t{\xdef\secsym{\the\secno.}\global\meqno=1\bigbreak\bigskip}
\def\sequentialequations{\def\eqnres@t{\bigbreak}}\xdef\secsym{}
\global\newcount\subsecno \global\subsecno=0
\def\subsec#1{\global\advance\subsecno by1\message{(\secsym\the\subsecno. #1)}
\ifnum\lastpenalty>9000\else\bigbreak\fi
\noindent{\it\secsym\the\subsecno. #1}\writetoca{\string\quad
{\secsym\the\subsecno.} {#1}}\par\nobreak\medskip\nobreak}
\def\appendix#1#2{\global\meqno=1\global\subsecno=0\xdef\secsym{\hbox{#1.}}
\bigbreak\bigskip\noindent{\bf Appendix #1. #2}\message{(#1. #2)}
\writetoca{Appendix {#1.} {#2}}\par\nobreak\medskip\nobreak}
%
%
\def\eqnn#1{\xdef #1{(\secsym\the\meqno)}\writedef{#1\leftbracket#1}%
\global\advance\meqno by1\wrlabeL#1}
\def\eqna#1{\xdef #1##1{\hbox{$(\secsym\the\meqno##1)$}}
\writedef{#1\numbersign1\leftbracket#1{\numbersign1}}%
\global\advance\meqno by1\wrlabeL{#1$\{\}$}}
\def\eqn#1#2{\xdef #1{(\secsym\the\meqno)}\writedef{#1\leftbracket#1}%
\global\advance\meqno by1$$#2\eqno#1\eqlabeL#1$$}
%
\newskip\footskip\footskip14pt plus 1pt minus 1pt 
\def\footnotefont{\ninepoint}\def\f@t#1{\footnotefont #1\@foot}
\def\f@@t{\baselineskip\footskip\bgroup\footnotefont\aftergroup\@foot\let\next}
\setbox\strutbox=\hbox{\vrule height9.5pt depth4.5pt width0pt}
\global\newcount\ftno \global\ftno=0
\def\foot{\global\advance\ftno by1\footnote{$^{\the\ftno}$}}
%
\newwrite\ftfile
\def\footend{\def\foot{\global\advance\ftno by1\chardef\wfile=\ftfile
$^{\the\ftno}$\ifnum\ftno=1\immediate\openout\ftfile=foots.tmp\fi%
\immediate\write\ftfile{\noexpand\smallskip%
\noexpand\item{f\the\ftno:\ }\pctsign}\findarg}%
\def\footatend{\vfill\eject\immediate\closeout\ftfile{\parindent=20pt
\centerline{\bf Footnotes}\nobreak\bigskip\input foots.tmp }}}
\def\footatend{}
%
%
\global\newcount\refno \global\refno=1
\newwrite\rfile
%
\def\ref{[\the\refno]\nref}%
\def\nref#1{\xdef#1{[\the\refno]}\writedef{#1\leftbracket#1}%
\ifnum\refno=1\immediate\openout\rfile=refs.tmp\fi%
\global\advance\refno by1\chardef\wfile=\rfile\immediate%
\write\rfile{\noexpand\Item{#1}\reflabeL{#1\hskip.31in}\pctsign}\findarg\hskip10.0pt}%
\def\findarg#1#{\begingroup\obeylines\newlinechar=`\^^M\pass@rg}
{\obeylines\gdef\pass@rg#1{\writ@line\relax #1^^M\hbox{}^^M}%
\gdef\writ@line#1^^M{\expandafter\toks0\expandafter{\striprel@x #1}%
\edef\next{\the\toks0}\ifx\next\em@rk\let\next=\endgroup\else\ifx\next\empty%
\else\immediate\write\wfile{\the\toks0}\fi\let\next=\writ@line\fi\next\relax}}
\def\striprel@x#1{} \def\em@rk{\hbox{}}
\def\lref{\begingroup\obeylines\lr@f}
\def\lr@f#1#2{\gdef#1{\ref#1{#2}}\endgroup\unskip}
\def\semi{;\hfil\break}
\def\addref#1{\immediate\write\rfile{\noexpand\item{}#1}} 
\def\listrefs{\footatend\vfill\supereject\immediate\closeout\rfile\writestoppt
\baselineskip=14pt\centerline{{\bf References}}\bigskip{\frenchspacing%
\parindent=20pt\escapechar=` \input refs.tmp\vfill\eject}\nonfrenchspacing}
\def\startrefs#1{\immediate\openout\rfile=refs.tmp\refno=#1}
\def\xref{\expandafter\xr@f}\def\xr@f[#1]{#1}
\def\refs#1{\count255=1[\r@fs #1{\hbox{}}]}
\def\r@fs#1{\ifx\und@fined#1\message{reflabel \string#1 is undefined.}%
\nref#1{need to supply reference \string#1.}\fi%
\vphantom{\hphantom{#1}}\edef\next{#1}\ifx\next\em@rk\def\next{}%
\else\ifx\next#1\ifodd\count255\relax\xref#1\count255=0\fi%
\else#1\count255=1\fi\let\next=\r@fs\fi\next}
\def\figures{\centerline{{\bf Figure Captions}}\medskip\parindent=40pt%
\def\fig##1##2{\medskip\item{Fig.~##1.  }##2}}
%
\newwrite\ffile\global\newcount\figno \global\figno=1
\def\fig{fig.~\the\figno\nfig}
\def\nfig#1{\xdef#1{fig.~\the\figno}%
\writedef{#1\leftbracket fig.\noexpand~\the\figno}%
\ifnum\figno=1\immediate\openout\ffile=figs.tmp\fi\chardef\wfile=\ffile%
\immediate\write\ffile{\noexpand\medskip\noexpand\item{Fig.\ \the\figno. }
\reflabeL{#1\hskip.55in}\pctsign}\global\advance\figno by1\findarg}
\def\listfigs{\vfill\eject\immediate\closeout\ffile{\parindent40pt
\baselineskip14pt\centerline{{\bf Figure Captions}}\nobreak\medskip
\escapechar=` \input figs.tmp\vfill\eject}}
\def\xfig{\expandafter\xf@g}\def\xf@g fig.\penalty\@M\ {}
\def\figs#1{figs.~\f@gs #1{\hbox{}}}
\def\f@gs#1{\edef\next{#1}\ifx\next\em@rk\def\next{}\else
\ifx\next#1\xfig #1\else#1\fi\let\next=\f@gs\fi\next}
\newwrite\lfile
{\escapechar-1\xdef\pctsign{\string\%}\xdef\leftbracket{\string\{}
\xdef\rightbracket{\string\}}\xdef\numbersign{\string\#}}
\def\writedefs{\immediate\openout\lfile=labeldefs.tmp \def\writedef##1{%
\immediate\write\lfile{\string\def\string##1\rightbracket}}}
\def\writestop{\def\writestoppt{\immediate\write\lfile{\string\pageno%
\the\pageno\string\startrefs\leftbracket\the\refno\rightbracket%
\string\def\string\secsym\leftbracket\secsym\rightbracket%
\string\secno\the\secno\string\meqno\the\meqno}\immediate\closeout\lfile}}
\def\writestoppt{}\def\writedef#1{}
\def\seclab#1{\xdef #1{\the\secno}\writedef{#1\leftbracket#1}\wrlabeL{#1=#1}}
\def\subseclab#1{\xdef #1{\secsym\the\subsecno}%
\writedef{#1\leftbracket#1}\wrlabeL{#1=#1}}
\newwrite\tfile \def\writetoca#1{}
\def\leaderfill{\leaders\hbox to 1em{\hss.\hss}\hfill}
\def\writetoc{\immediate\openout\tfile=toc.tmp
   \def\writetoca##1{{\edef\next{\write\tfile{\noindent ##1
   \string\leaderfill {\noexpand\number\pageno} \par}}\next}}}
\def\listtoc{\centerline{\bf Contents}\nobreak\medskip{\baselineskip=12pt
 \parskip=0pt\catcode`\@=11 \input toc.tex \catcode`\@=12 \bigbreak\bigskip}}
\catcode`\@=12 

\def\uln#1{\underline{#1}}

\def\plpl{{+\!\!\!\!\!{\hskip 0.009in}{\raise -1.0pt\hbox{$_+$}}
{\hskip 0.0008in}}}
\def\mimi{{-\!\!\!\!\!{\hskip 0.009in}{\raise -1.0pt\hbox{$_-$}}
{\hskip 0.0008in}}}

\def\items#1{\\ \item{[#1]}}
\def\ul{\underline}
\def\un{\underline}
\def\kd#1#2{\d\du{#1}{#2}}

\doit1{
\def\E{{\cal E}}
\def\Re{{\cal R}e\,}
\def\Im{{\cal I}m\,}
\def\order#1#2{{\cal O}({#1}^{#2})}
\def\rmp#1#2#3{Rev.~Mod.~Phys.~{\bf {#1}} (19{#2}) {#3}}
}

\def\framing#1{\doit{#1}
{\framingfonts{#1}
\border\headpic
}}

\framing{0}

{}~~~
\vskip 0.07in

{\hbox to\hsize{April, 1995\hfill UMDEPP 95--111}}
\par

\hsize=6.5in
\textwidth=6.5in

\begin{center}
\vglue 0.1in

{\large\bf Stationary Axisymmetric Black Holes,} \\
{\large\bf $N=2$ Superstring, and Self--Dual Gauge or Gravity
Fields}$\,$\footnote{This work is supported in part by NSF grant \#
PHY-93-41926.
}
\\[.1in]

\baselineskip 10pt

\vskip 0.18in

\doit1{
Hitoshi ~N{\small ISHINO}\footnote{E-mail: nishino@umdhep.umd.edu.} \\[.25in]
{\it Department of Physics} \\ [.015in]
{\it University of Maryland at College Park}\\ [.015in]
{\it College Park, MD 20742-4111, USA} \\[.12in]
and \\[.12in]
{\it Department of Physics and Astronomy} \\[.015in]
{\it Howard University} \\[.015in]
{\it Washington, D.C. 20059, USA} \\[.18in]
}

\vskip 0.3in

{\bf Abstract} \\[.1in]
\end{center}

\begin{quotation}

{}~~~We present interesting relationship between what are called stationary
axisymmetric black hole solutions for the vacuum Einstein equations in the
ordinary four-dimensions and exact solutions for self-dual
Yang-Mills fields in
flat $2+2$ dimensions which are nothing but the consistent backgrounds for
$~N=2$~ open superstring.  We show that any stationary axisymmetric
black hole solution for the former automatically provides an exact
solution for the latter.
We also give a nice relation between the physical parameters of
black holes and an invariant integral analogous to instanton charges for such a
self-dual Yang-Mills solution.  This result indicates that any general
black hole solution in the usual four-dimensions can be the
background of $~N=2$~ superstring at the same time.  We also give an
interesting
embedding of stationary axisymmetric solutions into the background of
$~N=2$~ closed superstring.   Finally we show some indication that the
Kerr solution with naked singularity (for $a > m$) is nothing else than the
gravitational field generated by a closed string.

\endtitle

\oddsidemargin=0.03in
\evensidemargin=0.01in
\hsize=6.5in
\textwidth=6.5in
\baselineskip 17pt

\centerline{\bf 1.~~Introduction}

The physical as well as mathematical significance of $~N=2$~ superstring theory
\ref\ntwo{L.~Brink and J.H.~Schwarz, \pl{121}{77}{185}.} manifests itself
in various contexts.  First of all, the consistent background for this theory
is
supposed to be either self-dual Yang-Mills (SDYM) field for {\it open} $~N=2$~
superstring or self-dual gravity for {\it closed} $~N=2$~
superstring \ref\ov{H.~Ooguri and C.~Vafa, \mpl{5}{90}{1389};
\np{361}{91}{469}; \ibid{367}{91}{83}.}\ref\ng{H.~Nishino and S.J.~Gates, Jr.,
\mpl{7}{92}{2543}.}.  The physical importance of the SDYM field comes from the
mathematical conjecture \ref\atiyah{M.~Atiyah, {\it unpublished}; R.S.~Ward,
Phil.~Trans.~R.~London {\bf A315} (1985) 51; N.J.~Hitchin, Proc.~London
Math.~IHE {\bf 55} (1987) 59.}  that the SDYM theory
may be the underlying theory of {\it all} lower-dimensional integrable
theories,
which are extremely important for many physical models.  In a recent series of
papers \ref\kng{S.J.~Gates, Jr.~and H.~Nishino, \pl{299}{93}{255};
S.V.~Ketov, H.~Nishino and S.J.~Gates, Jr., \np{393}{93}{149};
H.~Nishino, \pl{316}{93}{298}; \pl{324}{94}{315}.}\ref\nishinowznw{H.~Nishino,
\pl{316}{93}{298}.} we have also shown the generalization of this conjecture to
the case of self-dual {\it supersymmetric} YM and self-dual supergravity
(SDSG) theories.

It is then not surprising that some exact solutions for Einstein equation in
general relativity, which are also known to be or conjectured to be integrable
\ref\lk{See {\it e.g.}, M.~Lakshmanan and P.~Kaliappan, \jmp{24}{83}{795}.},
may
be embedded  into the SDYM theory in $2+2$ dimensions.  Independent of the
progress in this direction, it has been also recently pointed out that the
stationary axisymmetric solutions for vacuum Einstein equation is completely
separated by re-formulating the Ernst equation \ref\ernst{F.J.~Ernst,
\pr{167}{68}{1175}.} and its associated linear system in terms of a
non-autonomous Schlesinger-type dynamical system \ref\nicolai{D.~Korotkin and
H.~Nicolai, Hamburg University preprint (December, 1994).}     .

Independently of these developments, there has been also some progress
related to what is called Kerr solution \ref\kerr{R.P.~Kerr,
\prl{11}{63}{237}.}
with a ring singularity in axionic dilaton gravity may be regarded as a
soliton-like solution for the heterotic string theory \ref\sen{A.~Sen,
Tata preprint, TIFR-TH-92-57 (hep-th/9210050).}.  Interestingly it was
found that some limit of this solution near the singular ring coincides with
an exact solution for the heterotic string
\sen.  The importance of the Kerr solution strongly indicates the significant
relevance of more general axisymmetric black hole solutions to string
theories.

Based on these recent developments, we present in this paper explicit
relationships between what are called axisymmetric black hole exact solutions
in
general relativity in $1+3$ dimensions and the SDYM for the gauge group
$~GL(2,{\IR})$~ in $2+2$ dimensions which are nothing but the consistent
backgrounds for $~N=2$~ open superstring
\ov\ng\ref\siegelbackgrounds{W.~Siegel, \pr{47}{93}{2504}.}.
We show that any stationary
axisymmetric black hole solution in $1+3$ dimensions can be at the same
time an exact solution for SDYM fields in $2+2$ dimensions.  We also show a
nice
embedding of static axisymmetric stationary black hole solutions
into the backgrounds of $~N=2$~ {\it closed} superstring
\siegelbackgrounds.  We finally show that the
Kerr solution with the parameters $~a>m$~ has the energy-momentum tensor with
$~\d\-$function singularity
on a finite ring, indicating that this solution describes
the gravitational field generated by a closed string.

\newpage

\centerline{\bf 2.~~Embedding of Ernst Equation into SDYM}

We start with the review of what is called Ernst equation \ernst\ for
stationary
axisymmetric exact solutions for vacuum Einstein equations.  It is well-known
\ref\book{D.~Kramer, H.~Stephani, E.~Herlt and M.~MacCallum, {\it ``Exact
Solutions of Einstein's Field Equations''}, Cambridge University Press (1980).}
that any stationary axisymmetric space-time
$$d s^2 = f^{-1} \left[ e^{2 k} (d
z^2 + d\r^2 ) + \r^2 d\phi^2  \right]  - f (d t + \omega d\phi )^2 ~~
\eqno(2.1) $$
can be determined by a complex Ernst potential
$~\E (z,\r)$~ satisfying
$$\li{ &\left[ \fracmm{\r( \E \E ^* )_z}{(\E+\E ^*) ^2}
\right]_z  + \left[ \fracmm{\r( \E \E ^* )_\r}{(\E+\E ^*) ^2}
\right]_\r = 0 ~~,
{}~~~~ \left[ \fracmm{i\r(\E- \E^* )_z}{(\E+\E^*) ^2} \right]_z  +
\left[ \fracmm{i\r(\E- \E^* )_\r}{(\E+\E^*) ^2} \right]_\r = 0 ~~,{~~~~}
&(2.2b) \cr
& \left[ \fracmm{i\r(\E^{*\,2}\E_z - \E^2\E^*_z)}{(\E+\E^*)^2}
\right]_z + \left[ \fracmm{i\r(\E^{*\,2}\E_\r - \E^2\E^*_\r) }{(\E+\E^*)^2}
\right]_\r = 0 ~~.
&(2.2b) \cr } $$
In our notation the suffices $~_{z}$~
or $~_{\r}$~ indicate the partial differentiations with respect to these
variables, and the stars are for complex conjugates.
Eq.~(2.2b) is a necessary condition of the first two, and if the stationary
axisymmetric solution is further specified to be {\it static}, the Ernst
potential becomes real and (2.2b) is redundant.

In order to embed the Ernst equations into the SDYM fields
in flat $2+2$ dimensions, we need to choose a convenient frame in the latter,
specified by the coordinates
$~(x^\m) = (z,\r,\zeta, \t)$~ and the $~4\times 4$~ flat metric
\ref\belavin{See {\it e.g.}, J.F.~Pleba{\~ n}ski, \jmp{23}{82}{1126};
J.~Gegenberg and A.~Das, \grg{16}{84}{817}.}
$$\left(\eta_{\m\n}\right) = \pmatrix{0 & \s_3 \cr  \s_3
& 0\cr } ~~,
\eqno(2.3) $$
where $~\s_3~$ is the third Pauli matrix.  This new flat
$2+2$ dimensional space-time is entirely different from
the original $1+3$ dimensions (2.1).  In the
$2+2$ dimensions, the self-duality (SD) of the YM field: $~F_{\m\n}
= (1/2) \e_{\m\n}{}^{\r\s} F_{\r\s}$~ is  equivalent to the set of three
equations\footnotew{It is to be stressed that  even though the first two
equations imply that the components of the YM gauge field for these
two-dimensional sub-manifolds are pure gauge, the gauge field may  {\it
not} be globally gauged away like monopoles with non-trivial topology.  We will
see explicit examples shortly.}
$$ F_{1 2}{}^I = 0 ~~, ~~~~ F_{3 4}{}^I = 0 ~~, ~~~~ F_{13}{}^I = F_{2
4}{}^I ~~.
\eqno(2.4) $$
The indices $~{\scst I,~J,~\cdots}$~ are for the adjoint
representation of a YM gauge group, which will be sometimes suppressed like
(2.5).

Motivated by the $~2\times 2$~ real matrix representation for the Ernst
equation
\book, we choose natural YM gauge group to be
$~GL(2,\IR)$, specifying the YM potential as
$$\eqalign{&A_1 = +Q \S_0 + \frac 1{\sqrt 2} S (\S_1+ \S_2) ~~, ~~~~
  A_2 = -P \S_0 - \frac1{\sqrt 2} R (\S_1+ \S_2) ~~, \cr
&A_3 =  + \Hat S \S_0 - \frac 1{\sqrt 2} \Hat Q (\S_1+ \S_2) ~~, ~~~~
  A_4 = - \Hat R \S_0 + \frac 1 {\sqrt 2} \Hat P (\S_1+ \S_2) ~~, \cr }
\eqno(2.5) $$
where $~P,~Q,~R,~S$~ are real function only of $~(z,\r)$, while
$~\Hat P,~\Hat Q,~\Hat R,~\Hat S$~ are real functions only of $~(\z,\t)$, and
$~\S_I~({\scst I~=~0,~\cdots,~3})$~ are the generators for $~GL(2,\IR)$:
$$\S_0 = \half {\scst\pmatrix{1 & 0 \cr 0 & 1 \cr}}~~, ~~~~
\S_1 = \half {\scst\pmatrix{0 &
1 \cr 1 & 0 \cr}}~~, ~~~~
\S_2 = \half {\scst\pmatrix{0 & 1 \cr -1 & 0 \cr}}~~, ~~~~
\S_3 = \half {\scst\pmatrix{1 & 0 \cr 0 & -1 \cr}}~~.
\eqno(2.6) $$
By this assignment, we are constructing the $2+2$ dimensions as a product of
two
manifolds with the coordinates $~(z,\r)$~ and $~(\z,\t)$.
We now see that the SD conditions (2.4) implies
$$ \li{&P_z + Q_\r = 0 ~~, ~~~~ R_z + S_\r = 0 ~~,
&(2.7a) \cr
& \Hat P_\z + \Hat Q_\t = 0 ~~, ~~~~ \Hat R_\z + \Hat S_\t = 0 ~~.
&(2.7b) \cr } $$
The simplest assignment of these
functions to embed the Ernst equations (2.2) is
$$\eqalign{& P = \fracmm{\r\left( \E \E^* \right)_z}{(\E+\E^*)^2} ~~, ~~~~
  Q = \fracmm{\r\left( \E \E^* \right)_\r}{(\E+\E^*)^2}~~, ~~~~
  R = \fracmm{i\r\left(\E- \E^* \right)_z}{(\E+\E^*)^2} ~~, ~~~~
  S = \fracmm{i\r\left(\E- \E^* \right)_\r}{(\E+\E^*)^2} ~~. \cr}
\eqno(2.8) $$
Relevantly it is convenient to define
$$ T = \fracmm{i \r \big( \E^{*\,2}\E_z -
\E^2\E^*_z) }{(\Tilde\E+\Tilde\E^*)^2}
  ~~, ~~~~
  U = \fracmm{i \r \big( \E^{*\,2}\E_\r - \E^2\E^*_\r)
}{(\Tilde\E+\Tilde\E^*)^2}~~.
\eqno(2.9) $$
We assign similarly $~\Hat P,~ \Hat Q,~\Hat R,~\Hat S,~\Hat
T,~\Hat U$~ with $~(z,\r)$~ replaced by $~(\z,\t)$~ and the {\it hatted}
Ernst potential as a function only of $~(\z,\t)$, {\it e.g.,}$~\Hat P \equiv \t
(\Hat\E \Hat\E^*)_\z/(\Hat\E+\Hat\E^*)^2$.  We have thus a parallel structure
between the two  sets of coordinates $~(z,\r)$~ and  $~(\z,\t)$, so that the
$2+2$ dimensions are now decomposed into two sets of Ernst equation systems
each
for two sub-dimensions of stationary axisymmetric solutions.  There are lots of
other equally simple assignments, but our choice here is such that an
invariant integral (3.1) will have a certain manifest global symmetry,
as we will see later.

We mention that any known instanton solutions for
the Euclidean SDYM theory can be ``Wick-rotated'' to this $2+2$ dimensions.
The BPST single $~SU(2)$~ instanton
solution \ref\instanton{ A.A.~Belavin, A.M,~Polyakov, A.A.~Schwarz and
Y.S.~Tyupkin, \pl{59}{75}{85}.} in $4+0$ dimensions with the instanton number
$~k=1$~ can be re-casted into our $2+2$
dimensions for the non-compact gauge group $~SL(2,\IR) \subset GL(2,\IR)$:
$$\eqalign{A_1 = & \, 2 \left[\, + \z \S_3 + \t (\S_1 - \S_2) \,\right] G ~~,
{}~~~~A_2 = 2 \left[\, +\t \S_3 - \z (\S_1+\S_2) \, \right] G ~~, \cr
A_3 = & \, 2 \left[\, - z \S_3 +  \r (\S_1 + \S_2)  \, \right] G ~~, ~~~~
   A_4 =  2 \left[ - \r \S_3 + z (\S_1 - \S_2) \, \right] G ~~, \cr }
\eqno(2.10) $$
where $~G\equiv 1/(x_\m x^\m + b^2)$~
with an arbitrary constant $~b$, and $~x_\m x^\m
= 2(z\z - \r\t)$~ with the metric (2.3).

\vfill\eject

We can superimpose this instanton solution onto the above SDYM fields
constructed from black hole solutions, as long as the commutators between them
vanish.  Moreover, if we have a new set of solutions for the SDYM in $2+2$
dimensions, we may  ``Wick rotate'' it to the Euclidean $4+0$ dimensions.

\bigskip\bigskip\bigskip

\centerline{\bf 3.~~Black Hole Parameters and Invariant Integral}

Once we have succeeded in embedding the stationary axisymmetric solutions into
SDYM in $2+2$ dimensions, we can consider
``instanton charge'' for a SDYM field.  Since our space-time and
the YM gauge group are non-compact, there is no strict concept such as the
topological instanton charge for the SDYM fields.  Nevertheless we still can
formally define an analogous invariant integral:
$$ C\equiv \fracmm1{64\pi^2} \int d^4 x \, \e^{\m\n\r\s} F_{\m\n}{}^I
F_{\r\s}{}^I~~,
\eqno(3.1) $$
as a direct analog of the usual instanton charge.
For examples below we restrict the YM group to be $~GL(2,\IR)$.

The integrand in (3.1) is a total-divergence, and the only
contributions come from the surface terms.  We can
evaluate this charge for any known exact solutions such as the Kerr solution
\kerr\ or Tomimatsu-Sato (TS) solutions
\ref\ts{A.~Tomimatu and H.~Sato, \ptp{50}{73}{95}.}.  To
this end, it is convenient to use the prolate
spherical coordinates $~(x,y)$:
$$ z \equiv \k x y~~, ~~~~ \r \equiv \k{\sqrt{(x^2-1)(1-y^2)}}~~.
\eqno(3.2) $$
The constant $~\k$~ has the physical dimension of length.  We
can now rewrite (3.1) in terms of the surface integrals:
$$ C[\E,\Hat\E]  = \fracmm 1 {8\pi^2} \left( X[P,Q]  \, X[\Hat R,\Hat S]
- X[R,S] \, X[\Hat P,\Hat Q] \right) ~~,
\eqno(3.3) $$
where\footnotew{We consider only the connected patch of each coordinate
system for the integral.}
$$\eqalign{ &X[P, Q] \equiv \int d x^1\int d x^2
(P_z + Q_\r) \cr  =
& \, -\k \int_{-1}^1  d y \left[ \lim_{x\rightarrow\infty} (x P)
\fracmm y{\sqrt{1-y^2}}  - \lim_{x\rightarrow 1} ({\sqrt{x-1}} \, P) \fracmm
{{\sqrt 2}y}{{\sqrt{1-y^2}}}  + \lim_{x\rightarrow\infty} (x Q) -
\lim_{x\rightarrow 1} Q  \right]  \cr
& - \k \int_1^\infty d x \left[ \left\{ \lim_{y \rightarrow 1}
({\sqrt{1-y}} \, P) - \lim_{y\rightarrow -1} ({\sqrt{1+y}} \, P) \right\}
\fracmm {{\sqrt2}x}{\sqrt{x^2-1}} - \lim_{y\rightarrow 1} Q
- \lim_{y\rightarrow -1} Q  \right] ~~. \cr }
\eqno(3.4) $$
Another quantity $~X[R,S]$~ is simply obtained by
replacing $~(P,Q)$~ in (3.4) by $~(R,S)$.  It is clear now that our
invariant integral is a product of surface integrals like monopole
charges living on the two-dimensional sub-manifolds.

We can develop a more convenient formula for $~X[P,Q]$,
when some asymptotic form of the Ernst potential is known.  Suppose at
$~x\approx \infty$~ we have
$$ \E \approx 1 + \fracmm{F(y)} x + \order x{-2} ~~,
\eqno(3.5) $$
with a complex function $~F(y)$~ of $~y$.
The first term of (3.5) is
general enough for the boundary condition $~\Re\E \rightarrow 1$~
for asymptotically flat solutions, because the Ernst equation
(2.2) is invariant under the shift of $~\E$~ by arbitrary purely imaginary
constant \book.  If we define $~G(y) \equiv \Re F(y),~H(y) \equiv \Im F(y)$,
these functions are actually determined by the Ernst eqs.~(2.7) themselves
at this order:
$$ G(y) = a \ln\left( \fracmm{1-y}{1+y} \right) + G_0 ~~,   ~~~~
H(y) = c \ln\left(\fracmm{1-y}{1+y} \right) + H_0  ~~,
\eqno(3.6) $$
where $~a,c,G_0,H_0$~ are constants.  The asymptotic form of  the
$~g\low{00}\-$component of the original line element (2.1) at
$z\approx 0$~ and $~\r\approx\infty$~ namely $~y\approx 0,~x\approx\infty$~
must
be   $$ g\low{00} = - f = - \Re\E \approx -1 -
\fracmm{G_0} x + \order x{-2} \approx - 1 + \fracmm{2 m} \r +
\order\r{-2} ~~.
\eqno(3.7) $$
to accord with the Newtonian approximation.  This requires the residue
$~G_0$~ to be
$$ G_0 = - \fracmm {2m}\k ~~.
\eqno(3.8) $$

The asymptotic forms of ~$P,~\cdots,~S$ can be also fixed.  Using (3.5), we can
easily see that
$$\li{ & P \approx -\fracmm 1{2x} {\sqrt{1-y^2}} \left[ y G -
(1-y^2) G_y \right] + \order x{-2} ~~, \cr
& Q \approx - \fracmm1{2x} (1-y^2) \left( G + y G_y \right)
+ \order x{-2} ~~,
&(3.9a) \cr
& R \approx + \fracmm 1 {2x} {\sqrt{1-y^2}} \left[ \,y H - (1-y^2)
H_y \right]  + \order x{-2} ~~, \cr
& S \approx +\fracmm1 {2x} (1-y^2) (H + y H_y) + \order x{-2} ~~.
&(3.9b) \cr } $$
Without much loss of generality\footnotew{This property seems common to
asymptotically free black hole solutions.} we can also assume that
$$ \lim_{x \rightarrow 1} ({\sqrt{x-1}} \, P) =
\lim_{y \rightarrow \pm 1} ({\sqrt{1 \mp y}} \, P) =
\lim_{x \rightarrow 1} Q = \lim_{y\rightarrow \pm 1} Q = 0 ~~,
\eqno(3.10) $$
and {\it idem}.~for $~R$~ and $~S$.  We now see that
only two terms $~\lim_{x\rightarrow \infty} (x P)$~ and
$~\lim_{x\rightarrow \infty} (x Q)$~ in (3.4) remain under (3.6) and
(3.9), therefore $~X$'s is completely determined by the residues
$$X[P,Q] = \fracmm \k 2 \int_{-1} ^1 d y~ G(y) = \k G_0 = - 2 m~~,
{}~~~~ X[R,S] = \fracmm \k 2 \int_{-1} ^1 d y~ H(y) = - \k H_0~~,
\eqno(3.11) $$
and we can easily estimates the invariant integral for arbitrary exact
solutions:
$$ C[\E, \Hat\E] = \fracmm{\k^2}{8\pi^2} \left(G_0 \Hat H_0 - H_0 \Hat
G_0 \right)~~.
\eqno(3.12) $$

As is easily seen, there are many options for the choices of
the Ernst potentials $~\E$~ and $~\Hat\E$.  For example, we can choose
$~\E$~ to be the Kerr solution, while $~\Hat\E$~ to be
a TS solution.  We show below relevant residues needed for
representative cases of the Kerr solution with $~\d=1$~
\kerr\ (${\rm Kerr}^{\d = 1}$~ for short), Weyl solution (${\rm Weyl}^{\d}$),
and the $~\d=2$~ case of the TS solution \ts\ (${\rm TS}^{\d=2}$):

\noindent (i) Kerr Solution of $~\d=1$:
$$ \eqalign{&\E = \fracmm {\a-\b}{\a+\b} ~~,  ~~~~
\a\equiv p x - i q y ~~, ~~~~ \b\equiv 1~~~~(p^2 + q^2 = 1) ~~,
\cr
& \k G_0 = -\fracmm{2\k} p = - 2m~~, ~~~~ \k H_0  = 0~~.  \cr }
\eqno(3.13) $$

\noindent (ii) Weyl solution with general $~\d$:
$$ \E = \fracmm{(x-1)^\d}{(x+1)^\d}~~,
{}~~~~ \k G_0  = - 2\d\k = - 2m ~~, ~~~~\k H_0   =0~~.
\eqno(3.14) $$

\noindent (iii) TS solution for $~\d=2$:
$$ \eqalign{&\E = \fracmm {\a-\b}{\a+\b} ~\, , \,~~
\a \equiv p^2 x^4 + q^2 y^4 - 1 - 2 i p q x y (x^2 - y^2) ~\, , \,~~
\b \equiv 2 p x (x^2 - 1) - 2 i q y (1-y^2) ~\, , \cr
& \k G_0 = - \fracmm{4\k} p = - 2m
{}~~,  ~~~~ \k H_0 =0 ~~~~(p^2+q^2 =1) ~~. \cr}
\eqno(3.15) $$
If we choose any of these solutions for $~\E$~ and $\Hat\E$,
the invariant integral (3.12) vanishes due to $~H_0=0$.  We will give
an example of non-vanishing cases shortly.

The solutions for the Ernst equation (2.2) have a global
$~SL(2,\IR)$~ symmetry \book:
$$ \Tilde\E = \fracmm{a\E - i b}{ic\E + d} ~~~~~(ad - bc = 1)~~,
\eqno(3.16) $$
where the constants $~a,~b,~c,~d$~ are real, and all the {\it tilded}
quantities in this section are after a global $~SL(2,\IR)$~ transformation.
The asymptotic condition $~\Re\Tilde\E \rightarrow 1~(x\rightarrow
\infty)$~ yields
$$ c^2 + d^2 = 1~~,
\eqno(3.17) $$
due to $~\E\rightarrow 1$~ by (3.5).  The transformation (3.16) is
rewritten in terms of $~P, ~\cdots, U$:
$$\eqalign{ & \Tilde P = (a d + b c) P + b d R - a c T~~, ~~~~  \Tilde Q = (a d
+ b c) Q + b d S - a c U ~~, \cr  &\Tilde R = 2 c d P + d^2 R - c^2 T ~~, ~~~~
\Tilde S = 2 c d Q + d^2 S - c^2 U ~~, \cr
&\Tilde T = - 2 a b P - b^2 R + a ^2 T ~~, ~~~~
\Tilde U = - 2 a b Q - b^2 S + a ^2 U ~~. \cr }
\eqno(3.18) $$
Under this global symmetry, the different components of the YM field
strength (2.5) are transformed into each other.  Applying this to
(3.9), we see that the residues transform as
$$\Tilde G_0 = (a d + b c) G_0 + (a c - b d ) H_0 ~~, ~~~~
\Tilde H_0 = - 2 c d G_0 + (d^2 - c^2) H_0 ~~.
\eqno(3.19) $$
Under (3.17), this linear transformation preserves
its determinant to be unity.  This signals the existence of
a bilinear form of $~G_0 \Hat H_0 - H_0 \Hat G_0$~
invariant under such transformations in the solution space, and this is exactly
the combination we had in our invariant integral (3.3).  In other words, our
embedding rule (2.5) was chosen in such a way that the
integral (3.1) or (3.12) is invariant under the global $~SL(2,\IR)$~
symmetry for asymptotically flat solutions.

The invariance of the integral (3.9) under the global $~SL(2,\IR)$~
symmetry can be utilized to test whether two exact solutions are linked to each
other under such global transformations.  For example, we can see if the
$~{\rm TS}^{\d=2}$~ solution is connected to the $~{\rm Weyl}^{\d=2}$~ solution
by comparing invariant integrals: $~C_1[\E_1,\Hat\E_1]$~ and
$~C_2[\E_2,\Hat\E_2]$~ where $~\E_1 = {\rm Weyl}^{\d=1}, ~{\Hat\E_1}
=\Tilde{\rm Weyl}^{\d=2},~\E_2 = {\rm Weyl}^{\d=1},~\Hat\E_2 = {\rm
TS}^{\d=2}$, where {\it tilde} implies the exact solutions after applying a
transformation (3.16).  This is because if $~{\rm TS}^{\d=2} = \Tilde{\rm
Weyl}^{\d=2}$~ under such a transformation, we will get $~C_1 = C_2$.  By
studying the values of parameters $~a,~b,~c,~d$~ that satisfy $~C_1 = C_2$,
we can see their connectedness for these special values.
In fact, we easy see that they are not really connected, because $~C_1 =
0,~C_2 = (2m^2 c d)/\pi^2 $, and the inspection of the possible
cases $~c=0~$ and $~d=0$~ reveals no connectedness.  Even though this
example is rather trivial, we can use this invariant integral to test
a ``newly'' found exact solution, to see its connectedness with any known
exact solution, whenever the direct comparison is difficult.

\bigskip\bigskip\bigskip

\centerline{\bf 4.~~Stationary Axisymmetric Black Holes for Closed
$~N=2$~ Superstring}

We have so far dealt with the SDYM fields for {\it open} $~N=2$~ superstring.
We may wonder about the SDSG for {\it closed} $~N=2$~ superstring.
In fact, we have already presented such embedding of the dilaton
black hole solution \ref\dilatonbh{E.~Witten, \pr{44}{91}{314}; G.~Mandal,
A.M.~Sengupta and S.R.~Wadia, \mpl{6}{91}{1685}; T.~Eguchi,
\mpl{7}{92}{85}.} into the backgrounds \siegelbackgrounds\ of {\it closed}
$~N=2$~ superstring \ref\nishinobh{H.~Nishino,
\pl{324}{94}{315}.}.  Below we will show how the embedding of stationary
axisymmetric vacuum solutions can be also embedded into such backgrounds.

The consistent background fields for closed $~N=2$~ superstring form a
multiplet
of $~N=8~$ SDSG in $2+2$ dimensions \siegelbackgrounds.  Among bosonic
background fields, we require that the fields $\phi_{A B C D},~ A_{\m}{}^{A
B},~
B_{\m}{}^{A B}$~ \siegelbackgrounds\footnotew{We use the same notation as in
ref.~\nishinobh.  Consistency of such truncation can be confirmed by studying
the original set of all the field equations \siegelbackgrounds, which we skip
in
this paper.} are all zero for simplicity in our prescription.
Eventually our non-vanishing fields are $~(e\du\m m, \o\du\m {r
s})$,\footnotew{The indices  $~{\scst \m,~\n,~\cdots}$~ and $~{\scst
m,~n,~\cdots}$~ are respectively the  curved and local Lorentz coordinates in
$2+2$ dimensions with the coordinates $~(x^\m) = (z,\r,\z,\t)$.} where
$~\o\du\m{r s}$ is self-dual for the indices $~^{r s}$, and the relevant
field equations are
$$\li{&2\e^{ \m \n \r \s} {  D}_{ \n}  {\L}  _{ \r \s m} - \e\, ^{ \m \n
\r \s} \L\du{ \n m} n   T_{ \r \s n} = 0 ~~, {\hskip 0.6in}
&(4.1a) \cr
& \e\,^{ \m \n \r \s} \left[\,   e\du{ \n}{ \[  m |}
  \,  T\du{ \r \s}{|  n\]}
  - \half  \e\, \ud{ m n}{ r s}   e\du\n r
  T\du{ \r \s} s \, \right] =0 ~~.
&(4.1b) \cr } $$
The torsion tensor is $~T\du{\m\n} m \equiv
2(\partial_{[\m} e\du{\n]} m + \o\du{[\m}{m n} e_{\n] n})$.

As the simplest solutions for (4.1) we have $~\o\du\m{r s}= 0$~
implying the lack of local Lorentz symmetry in the system,
while the vierbein is specified as
$$(e\du\m m) = \pmatrix{ 1-Q & S & 0 & 0 \cr
P & 1 - R & 0 & 0 \cr
0 & 0 & 1 - \Hat S &  \Hat Q \cr
0 & 0 & \Hat R & 1 - \Hat P \cr }~~,
\eqno(4.2) $$
where we use (2.8), the flat metric $~\eta_{m n}$~ of (2.3), and the
previous rule for {\it hatted} fields.
It is now straightforward to show that (4.2) yields ~$T\du{\m\n} r = 0
{}~$ similar to the previous case of SDYM in the flat space-time, and
thereby all the field equations in (4.1) are satisfied.

This torsion is like an analog of the SDYM field, and we expect
an invariant integral
$$ C' \equiv c \int d^4 x
{}~\e^{\m\n\r\s} T\du{\m\n} m T_{\r\s m}
= 8c \left( X[P,Q] X[\Hat R,\Hat S] - X[R,S] X[\Hat P, \Hat Q] \right) ~~,
\eqno(4.3) $$
which exactly coincides with (3.3) up to a constant:
$~C'=\hbox{const.}\,\times C$.
Even though the vierbein with vanishing anholonomy coefficients seems trivial,
non-vanishing integral (4.3) indicates the non-trivial feature of this
gravitational system with the connection fields that can
not be globally gauged away.  Like the previous SDYM, the embedding
(4.2) is fixed in such a way that $~C'$~ is invariant under the global
$~SL(2,\IR)$~ group for asymptotically flat solutions, which can be used as an
index to see the link between solutions.

The above example seems the simplest, but there may be other
non-trivial embedding scenarios we can develop with more non-trivial
vierbeins.  In any case, it is now obvious that any stationary axisymmetric
black  hole solution in $1+3$ dimensions can be also a consistent background in
$2+2$ dimensions for the closed as well as open
$~N=2 $~ superstring at the same time!

\newpage

\centerline{\bf 5.~~Kerr Solution as Gravity around Closed String}

Our results so far can be also reinterpreted based on the philosophy that the
$~N=2$~ superstring is expected to be ``Master Theory'' \ng\kng\ of all the
other superstring theories.  We have seen black
hole solutions for general relativity (the backgrounds of $~N=1$~ superstring
in $1+3$ dimensions) are embedded into the background field equations of
$~N=2$~
superstring in $2+2$ dimensions.  There is other supporting evidence for the
$~N=2$~ superstring to be the Master Theory of $~N\le 1$~ (super)string
theories, {\it e.g.,} we have shown \nishinowznw\ that
$~N=1$~ superconformal Wess-Zumino-Novikov-Witten sigma-models themselves are
embedded into the SDYM fields as the consistent backgrounds of
$~N=2$~ superstring, indicating that $~N=1$~ superstrings
exist as the direct target space of $~N=2$~ superstrings.
If this is indeed the case, we can suspect that the black hole solutions
in $1+3$ dimensions are nothing but the gravitational fields generated by
$~N\le 1$~ (super)strings themselves.\footnotew{This conjecture appears
in many different contexts \sen\ref\strominger{See, {\it e.g.},
G.T.~Horowitz, in Proceedings
{\it ``String Theory and Quantum Gravity '92''} (Trieste 1992);
A.~Strominger, in Proceeding of Les Houches Summer School,
(Les Houches, Aug.~1994); {\it and references therein}.},
but our motivation is stronger by our philosophy
based on $~N=2$~ superstring supported also by mathematics.}

Motivated by this observation, we try below to show that the
the matter energy density for the Kerr solution \kerr\ with $~a>m$~
has a manifest naked $~\d\-$function singularity
at a finite radius on the equatorial plane.  We start with
the Kerr solution with the
oblate spherical coordinates $~(u,y)$:
$$ \li{& d s^2 = m^2\left[ - e^{2\n} d t^2 + e^{2\psi} (d\varphi
- \Omega d t)^2 + e^{2\m\low 2} d u^2 + e^{2\m\low 3} d y^2 \right] ~~,
& (5.1) \cr
&\Omega = \fracmm{2 q^2 (\Hat p u +1)}{m D} ~~, ~~~~ e^{2\n} =
\fracmm{\Hat p^{\,2} (u^2+1) B} D ~~, ~~~~ e^{2\psi} = \fracmm{(1-y^2) D}
B~~, \cr
& e^{2\m\low 2} = \fracmm B{u^2+1}~~, ~~~~ e^{2\m\low 3} = \fracmm B
{1-y^2} ~~,  \cr
& B \equiv (\Hat p u+1)^2 + q^2 y^2~~, ~~~~
D \equiv \left[ (\Hat p u +1)^2 + q^2
\right]^2 - \Hat p^{\,2} q^2 (u^2 + 1) (1-y^2)~~,
& (5.2) \cr }  $$
where $~\n,~\m\low 2,~\m\low 3,~\psi$~ are functions only of $~(u,y)$,
and $~q^2 = 1 + \Hat p{\,}^2$.  We use the Kerr solution for the case $~a >
m$~ analytically continued from the case $~a\le m$~ with the
prolate coordinates with $~p^2 + q^2 = 1$.  Now the singularity is
{\it naked} in this Kerr solution and hence it really ``exists'' on the
ring $~\Hat p u + 1 = 0,~y=0$, where the Riemann tensor diverges.  Following
ref.~\ref\bardeen{J.M.~Bardeen, Astrophys.~Jour.~{\bf 162}
(1970) 71.}, the $~00\-$component of the matter energy-momentum tensor is
computed {\it via} the Einstein tensor in the ``locally non-rotating reference
frame'' as
$$\eqalign{ & 8\pi T_{(0)(0)} = R_{(0)(0)}
+ \half R =  m^{-2} e^{-\psi- \m\low 2-\m\low 3}
\left\{ \left[ e^{\m\low 3-\m\low 2} (e^\psi)_u \right]_u
- \left[ e^{\m\low 2-\m\low 3} (e^\psi)_y \right]_y \right\} \cr
& + m^{-2} e^{-\m\low 2 -\m\low 3} \left\{ \left[ e^{-\m\low 2} (e^{\m\low 3})
_u \right]_u - \left[ e^{-\m\low 3} (e^{\m\low 2})_y \right]_y
\right\} + \fracm14 m^{-2} e^{2\psi- 2\n} \left[ \Omega_u^2 e^{-2\m\low 2}
- \Omega_y^2 e^{-2\m\low 3} \right] ~~. \cr } {~~}
\eqno(5.3) $$
where the index $~_{(0)}$~ denotes the locally non-rotating $~0\-$th
coordinates, and subscripts $~_{u,~y}$~ are for partial derivatives.  Since the
Kerr metric is a vacuum solution, (5.3) vanishes everywhere {\it except} the
above naked ring singularity.

Our purpose is to show that this energy-momentum tensor component has a
$~\d\-$function at the naked singularity, like
$~\d(\Hat p u+1) \d(y)$, because the energy-momentum tensor vanishes everywhere
else as a vacuum solution, but at the same time its $~u$~ and $~y\-$integral
should give a non-vanishing mass.  Showing this is generally
difficult, because unless we have an appropriate ``regulator'' function, we
will simply get a vanishing result.  There have been also some
trials such as disk like
mass distributions \ref\israel{See {\it e.g.}, W.~Israel, \pr{2}{70}{641}.}
to explain the singularity, but none of them seems
to give the {\it manifest} $~\d\-$function singularity.  Our approach is
also different from the interpretation by complex hyperbolic string
in ref.~\ref\brinskii{A.~Ya Brinskii, \pla{185} {94}{441}; Moscow
preprint (hep-th/9503094).}.

A similar regulator is found in particle physics for a Coulomb potential
$~\phi(r)$~ for the Laplace equation, which can be regularized by the
Yukawa potential in three dimensions:
$$ \Delta \phi (r) = \lim_{\b\rightarrow 0}
\left( \fracmm{d^2}{d r^2} + \fracmm 2 r
\fracmm{d}{d r}\right) \fracmm {e^{-\b r}} {4\pi r} =
\lim_{\b\rightarrow 0} \left(\b^2 \, \fracmm {e^{-\b r}} {4\pi r}\right) ~~,
\eqno(5.4) $$
yielding
$$ \int_0^\infty d r ~4\pi r^2 ~\Delta \phi (r) = \lim_{\b\rightarrow 0} \b^2
\int _0^\infty d r \left( r e^{-\b r}\right) = +1 ~~,
\eqno(5.5) $$
implying that the r.h.s.~of (5.4) is actually a $~\d\-$function: $~
\d(r)/(4\pi r^2)$~ instead of $~0$.  If we take the $~\b\rightarrow 0$~ limit
{\it before} the $~r\-$integral, we get simply zero with no
singularity.  This fundamental feature can be understood
easily, but it is generally difficult to find a manifest regulator to
show the singularity, especially when the system is non-linear.\footnotew{We
could easily rely on Fourier transforms or test functions, if the system
were linear.}  Here we show an explicit working-example of such a regulator.

Our regulator acts only on the $~g_{u u}$~ and $~g_{y y}\-$components of the
metric:
$$ e^{\Tilde\m\low 2} = S(r) e^{\m\low 2}~~, ~~~~e^{\Tilde \m\low 3} = S(r)
e^{\m\low 3}~~ , ~~~~ S(r) \equiv 1 + b \, \a e^{-\a r/m}~~.
\eqno(5.6) $$
where the {\it tilded} quantities are regularized ones, and $~S(r)$~ is
our regulator, which goes to unity as the parameter $~\a\rightarrow 0$.
The real constant $~b$~ will be fixed shortly, and $~r$~
is the Boyer-Lindquist radial coordinate: $~r\equiv m(p x + 1)$~ or $~r
\equiv m(\Hat p u+1)$.  This $~S(r)$~ is an analog of $~\exp(- \b r)$~ in
(5.5).  Due to the cancellation in the exponents and $~y\-$independence of
$~S(r)$, it is only the third term in (5.3) that will contribute.

We first fix the constant $~b$~ in (5.6), applying it to the Schwarzschild
metric.  The spacial volume integral of (5.3) yields the
non-zero result:
$$\lim_{\a\rightarrow 0} \int_0^\infty d\xi \int_{-1} ^1 d y
\int_0^{2\pi} d\varphi {\sqrt{-\Tilde g}} \, m \, \left.\Tilde T^{0 0}
\right|_{\rm Schwarzschild}
= \half b \, m ~~.
\eqno(5.7)  $$
where $~\xi\equiv r/m = x+1$.  This implies the existence of the
$~\d\-$function singularity
$$m^2 \left. T^{0 0}\right|_{\rm Schwarzschild}
= \fracmm {b m} {8\pi r^2}\d(r)~~.
\eqno(5.8) $$
Due to the original dimensionless time coordinate, $~m^2T^{0 0}$~
corresponds to the energy density.
To accord with the standard point mass distribution \ref\mtw{{\it See, e.g.,}
C.W.~Misner, K.S.~Thorne and J.A.~Wheeler, {\it ``Gravitation''} (Cambridge,
Mass.~1972); R.M.~Wald, {\it ``General Relativity''} (Univ.~of Chicago Press,
1984).}, we fix
$$b = 2 ~~.
\eqno(5.9) $$

We mention the following important points to simplify the computation of
extracting only $~\order\a 0$ part of the total expression in (5.7):

\noindent (i)~~We perform a double-expansion for the integrand:
$~\sum_{k,n} A_{k,n}(y) \a^k ~(\xi+1)^n \exp\,(-l\a\xi)$~
\newline
where the integer $~l~(l\ge1)$~ is fixed, and
the $~\exp(-l \a\xi)$~ is {\it not} expanded.

\noindent (ii)~~The relevant terms can be estimated by the simple
integrals of the type
\newline
$\a^k \int_0^\infty d \xi~(\xi+1)^n \exp\,(-l\a\xi)$ with $~k\ge 1,~n\le 2$.

Some remarks are in order:  For (i) we can expand around any negative value of
$~\xi$~ instead of $~-1$, but the result does not depend on such a value: It is
just to avoid the singularity at $~\xi=0$.  For
(ii) all such integrals for $~n\le -2$~ with $~\a^k~(k\ge 1)$~ in front are
easily shown to vanish, while the subtle case $~k\ge 1,~n=-1$~ can be also
shown
to be of the order $~\a^k\ln\left[(\a+1)/\a \right]$~ which again vanishes as
$~\a\rightarrow 0$.  The only non-zero contributions in the
above computation are the cases $~(k,n)=(1,0),~(2,1)$~
and $~(3,2)$, which are all finite.

Following these points, we apply our regulator (5.6) now to the Kerr
solution (5.2):\footnotew{The lower limit of the $~\xi$-integral
can be any finite non-positive number, as long as the integral includes the
singularity at $~\xi=0$.}
$$ \lim_{\a\rightarrow0} \fracmm1{\Hat p}\int_0^\infty d\xi\int_{-1}^1 d y
\int_0^{2\pi} d\varphi ~{\sqrt{-\Tilde g}}\, m \, \left. \Tilde T^{0 0}
\right|_{\rm Kerr} = m ~~,
\eqno(5.10) $$
where now $~\xi\equiv r/m = \Hat p u + 1$.  This accords with the asymptotic
total mass of the system \mtw.
Since the whole integrand vanishes {\it except} the singularity as the vacuum
Kerr solution as $~\a\rightarrow 0$,
this implies the existence of a $~\d\-$function singularity:
$$ m^2 \left. T^{0 0}\right|_{\rm Kerr}
= \fracmm{\Hat p}{2\pi{\sqrt{-g}}} \d(\xi) \d(y) =
\fracmm {m^3} {2\pi a^2 (\r-a)^2} \d(\r - a) \d(\cos\theta) ~~,
\eqno(5.11) $$
in terms of the coordinates $~z\equiv m\Hat p u y,~\r \equiv m\Hat p
{\sqrt{(u^2+1)(1-y^2)}}$.  This means the existence of such
ring-like mass distribution around the circle at the finite
radius of $~\r = a$~ on the equatorial plane $~\theta=\pi/2$,
which is nothing but a ``closed string''!  It is interesting that
we have obtained the normalized total mass $~m$~ also for the Kerr solution,
providing another support for the validity of our regulator (5.6).  It seems
that our regulator can be applied also to other axisymmetric
solutions.

\newpage

\centerline{\bf 6.~~Concluding Remarks}

In this paper we have presented an interesting relationship between what are
called stationary axisymmetric black hole solutions and SDYM fields in
$2+2$ dimensions, which are nothing but the consistent backgrounds for $~N=2$~
{\it open} superstring.  We have confirmed that any
stationary axisymmetric solution of the vacuum Einstein equation can be the
backgrounds of $~N=2$~ open superstring at the same time!  We have found
interesting relations between the parameters for the axisymmetric black holes
and the invariant integrals analogous to topological instanton charges.  In
particular, the invariant integral of any axisymmetric black hole solution  is
determined by the residues in the asymptotic expansion of the Ernst  potential,
like asymptotic masses of the black holes.   We have also performed a nice
embedding of these black hole  solutions into the background vierbein of the
$~N=2$~ {\it closed} superstring, and developed a similar invariant integral.
We finally showed the indication that the Kerr solution for $~a>m$~ is
nothing but  the gravitational field around a closed string itself.

To our knowledge, our work is the first
attempt to relate the gravitational exact solutions to the backgrounds
for the $~N=2$~ superstring {\it via} SDYM and SDSG fields in an
explicit way.  As a by-product, our method also gives the
series of new exact solutions for the SDYM field by superpositions of
already-known instanton solutions such as the BPST solution onto SDYM fields
constructed from a black hole solution.

Our invariant integrals both for the SDYM and SDSG fields have
the manifest global $~SL(2,\IR)$~ symmetry acting on
asymptotically flat solutions in addition to the local $~GL(2,\IR)$~
gauge group.  Therefore we can also use
this integral as a convenient index indicating the
``connectedness'' of two given solutions under such global transformations.

The clear relationship between the stationary axisymmetric black holes and
$~N=2$~ superstring is also interesting, considering the
recent development about the Kerr solution in the axion dilaton gravity
coinciding with the background solutions for heterotic string \sen.  The
recent results relating conformal field theory and black hole
solutions \nicolai, or the interpretation of the ring singularity in the
Kerr solution interpreted as a complex hyperbolic string \brinskii, also
support this philosophy.  In contrast to the usual bosonic or
$~N=1$~ superstring theory accommodating black hole solutions
in $1+3$ dimensions \strominger, our result shows the direct
link between $~N=2$~ superstring in $2+2$ dimensions and black holes in $1+3$
dimensions.  We have thus provided another strong motivation of investigating
black hole solutions in general relativity based on superstring physics.

Our explicit results also strongly support the philosophy that the $~N=2$~
superstring theory is really the Master Theory of all (supersymmetric)
integrable systems in lower-dimensions \ng\kng.  It is reasonable that the
background field equations of such Master Theory are of the
first order, in particular the SD equations are the
most natural first-order equations for bosonic fields.  This is because
lower-order differential equations are more universal as embedding
equations in such a Master Theory.  In our examples, we
have seen that the black hole solutions for the second-order
Einstein field equation in the $1+3$ dimensions are embedded
into the first-order SDYM or SDSG equations for the
backgrounds of $~N=2$~ superstrings in $2+2$ dimensions.
Even though at first glance the embedding of exact solutions in $1+3$
dimensions
into $2+2$ dimensions is rather bizarre due to the
fundamental difference in signature, once we have understood the
significance of the integrable systems governing the basic differential
equations {\it via} self-dual theories, it becomes easy to comprehend the
naturalness of such embeddings.

Motivated by this philosophy, we conjecture that any black hole
solution in $1+3$ dimensions for general relativity is nothing but
the gravitational field generated by $~N\le 1$~ (super)strings.  This is
because
general relativity is the consistent background for $~N\le 1$~ superstring,
which in turn is to be the target space for the $~N=2$~ superstring
\nishinowznw.  In fact, we have shown that the $~00\-$component of the
energy-momentum tensor for the Kerr solution with $~a>m$~ actually has the
$~\d\-$funciton singularity on the ring at $~\r=a,~\theta=\pi/2$, using a
particular regulator for some metric components, which seems more applicable to
other solutions.  Our study provides strong
evidence that string physics is playing important roles also for
black hole physics.

Our results provide a completely new motivation for the observation of black
holes in real space.  So far string theory had been regarded as
physics at ``unrealistically'' high energy around the Planck mass that can
not to be easily realized by the present ``low-energy'' technology.
However, thanks to the encouraging results connecting
strings with black holes, string theory has now gained more realistic
aspects that can be probed by the exploration of black holes in
the real universe.

\bigskip\bigskip

The author is grateful to many experts in the fields, in particular D.~Brill,
T.~Jacobson and C.~Misner for helpful discussions and for information about
important references.  We are also indebted to S.J.~Gates, Jr.~for helpful
suggestions and encouragement.

\vfill\eject

\footatend\vfill\supereject\immediate\closeout\rfile\writestoppt
\baselineskip=14pt\centerline{{\bf References}}\bigskip{\frenchspacing%
\parindent=20pt\escapechar=` \input refs.tmp\vfill\eject}\nonfrenchspacing

\end{document}